\documentclass[prd,preprint,a4paper,superscriptaddress,nofootinbib,11pt]{revtex4}
\usepackage{amsmath,amsfonts,amssymb}
\usepackage{txfonts}
\usepackage{epsfig}
\usepackage{graphics}
\usepackage[T1]{fontenc}
\usepackage[safe]{textcomp}
\usepackage{dsfont}

\usepackage{csquotes}

\begin{document}

\begin{center}

{\bf  \Large Spectral Dimension of kappa-deformed space-time}
 
\bigskip

\bigskip

Anjana. V {\footnote{e-mail: anjanaganga@gmail.com}}and E. Harikumar  {\footnote{e-mail: harisp@uohyd.ernet.in }} \\
School of Physics, University of Hyderabad,
Central University P O, Hyderabad-500046,
India \\[3mm] 

\end{center}

\setcounter{page}{1}
\bigskip

 \begin{center}
  {\bf   Abstract}
  \end{center} 
  
  We investigate the spectral dimension of $\kappa$-space-time using the $\kappa$-deformed diffusion equation. The 
  deformed equation is constructed for two different choices of Laplacians in $n$-dimensional, $\kappa$-deformed 
  Euclidean space-time. We use an approach where the deformed Laplacians are expressed in the commutative space-time 
  itself. Using the perturbative solutions to  diffusion equations, we calculate the spectral dimension of 
  $\kappa$-deformed space-time and show that it decreases as the probe length decreases. By introducing a 
  bound on the deformation parameter, spectral dimension is guaranteed to be positive definite. We find that, for 
  one of the choices of the Laplacian, the non-commutative correction to the spectral dimension depends on the 
  topological dimension of the space-time whereas for the other, it is independent of the topological dimension.
  We have also analysed the dimensional flow for the case where the probe particle has a finite extension, unlike 
  a point particle.

\newpage
 
\section{Introduction}

Many different paradigms have been developed and employed to unravel the structure of space-time at
microscopic scale, in order to obtain the consistent quantum mechanical description of gravity. Dimensional flow
\cite{dm} is a characterising behaviour that is common to these approaches such as causal dynamical triangulations
\cite{dm}, asymptotically save gravity\cite{asg}, loop gravity\cite{loop}, deformed(or doubly) special relativity\cite{dsr}, non-commutative 
space-times\cite{nc1,nc2,nc3}, Horava-Lifschits gravity\cite{hl}, and relative-locality\cite{rl1,rl2,rl3,rl4}. 
All these approaches show that the effective dimensions felt by a fictitious test particle reduces at high energies
(i.e., when the length scale probed is very small). This variation of dimensions with the probe scale at high 
energies suggest the possibility of a fractal structure of the space-time at extremely short distances. This 
reduction of the effective dimensions is of important consequences as gravity is known to be renormalisable in 
two dimensions. 

Prediction of the existence of a fundamental length scale is another common trait shared by various frameworks 
developed to investigate microscopic theory of space-time. Non-commutative geometry provides an elegant mathematical 
framework to incorporate this fundamental length scale\cite{nc4}. In non-commutative setting, usual notions of 
space-time points get blurred and space-time becomes fuzzy. A particle undertaking a random walk in such a fuzzy 
space-time may not be able to access all the dimensions. Thus, the effective dimensions felt by this test particle 
can be different from the usual topological dimensions of the space-time. Spectral dimension has been used to study 
this change in the effective dimensions at high energies.

The usual notion of the dimension of space is that it is the exponent that characterising the change in the volume of an 
object with change in its size.
%\footnote{For a sphere, we know that the volume changes as the third power of its radius, showing the dimension of the space is three.}. 
Thus one defines the dimension as $d=LT_{size\to 0} 
\frac{log volme}{log size}$. Geometrical aspects of space at short distances are studied using the spectral 
dimension, which in the large length scale produces the same value as the usual dimension. Spectral dimension is
calculated from the solution to the diffusion equation defined in a Euclidean space with a given metric.  
The motion of the particle in such a space is governed by the diffusion equation
  \begin{equation}
   \frac{\partial }{\partial \sigma}U(x,y;\sigma) = \mathcal{L} U(x,y;\sigma)
  \end{equation}
 where $\sigma$ is fictitious diffusion time with dimension of length squared, $\mathcal{L}$ is the Laplacian 
and U(x,y;$\sigma$) is the probability density of the test particle to diffuse from the initial position $x$ 
to another point $y$, in diffusion time  $\sigma$. The spectral dimension of this space is related to the
trace of the diffusion probability known as return probability which measures the probability to find the particle 
returning  back to the starting point after a diffusion time $\sigma$. Thus, using the solution for the diffusion 
equation, the return probability is defined as
  \begin{equation}
   P(\sigma)= \frac{\int d^n x \sqrt{det g_{\mu \nu}}U(x,x;\sigma)}{\int d^n x \sqrt{det g_{\mu \nu}}}\label{rp}
  \end{equation}
where $g_{\mu \nu}$ is the metric of the underlying space. Spectral dimension $D_s$ is extracted by taking 
the logarithmic derivative of return probability P($\sigma$)
  \begin{equation}
  D_s = -2 \frac{\partial \ln P(\sigma)}{\partial \ln \sigma}.\label{specdim}
  \end{equation}
It is easy to see that for usual $n$-dimensional Euclidean space(where the metric $g_{\mu\nu}=\delta_{\mu\nu}$), 
the spectral dimension turns out to be $n$ itself. It is well known that the spectral dimension of quantum 
gravity models depends on the diffusion time $\sigma$ and it decreases smoothly to a lower value from four, 
at small values of $\sigma$ and this is known as dimensional flow\cite{dm}. It was shown that the spectral 
dimension reduces to $2$ in the UV (i.e., when probed at extremely short length scales) from $4$ at IR in 
many cases\cite{dm, asg,loop, hl} whereas, for  non-commutative space-times, for certain choices of the Laplacian,
the spectral dimension reduces to a lower value (but not $2$) and for other choices, one gets higher values for 
spectral dimension showing super diffusion\cite{dsr,nc1,nc2,nc3}. In recent times a novel approach to understand 
quantum gravity known as the principal of 
relative locality \cite{rl1,rl2,rl3,rl4} was introduced. In this approach locality of an event is 
also observer dependent as simultaneity in special and general relativity. In this approach  phase space 
is at a fundamental level than the space-time and argue that the phase space relevant for discussion of quantum 
gravity should have a non trivial geometry as first suggested in \cite{Born}.

Dimensional flow for a model compatible with deformed special relativity principle (DSR) has been analysed in \cite{dsr}. 
Starting with the metric in momentum space and with a specific choice of Laplacian (constructed using an invariant scalar), 
return probability in DSR compatible space-time was calculated. Using this, the spectral dimension was obtained and its 
variation with respect to parameters characterising the deviation of the Laplacian from the usual case was studied 
numerically. It was shown that the spectral dimension increases to $6$ in the high energy scale showing super diffusion 
and take the expected value of $4$ in the low energy scale. It was also noted that the spectral dimension do become 
non-integer for certain values of the parameters.

In \cite{nc1}, using numerical methods, fractal nature of the $\kappa$-Minkowski space-time was studied. Here, the 
spectral dimension is calculated from the return probability obtained using a specific form of $\kappa$-deformed
Laplacian, in the momentum space. Here, a modified integration measure, in the momentum space, which is invariant 
under the $\kappa$-deformed Lorentz algebra was also used.  After carrying out the integrations, numerically, 
it was shown that the spectral dimension reduced to three (3) in the limit of diffusion time going to zero. 
Further, using a different basis for the $\kappa$-deformed Lorentz algebra, same behaviour of spectral dimension 
was obtained, but with a different choice for the Laplacian. 

Variation of spectral dimension with the probe scale in a non-commutative 
space-time was analysed using another approach in \cite{nc2}. In this paper, the entire effect of non-commutativity 
was introduced through a modification of the initial condition satisfied by the solution to the 
diffusion equation. Since non-commutativity will lead to smearing of point objects, the initial condition 
was taken as a Gaussian(instead of a Dirac delta function) with width controlled by the minimal length 
introduced by the non-commutativity of the space-time.  In this case, the spectral dimension was shown 
to be a function of diffusion time and the minimal length parameter. In this case, at very high energies, 
space-time dimension reduces to zero. It was shown that in the limit where the diffusion time $(s)$ is of the 
same order as the (square of ) minimal length $(l_{min})$, the spectral dimension becomes $2$ , For 
trans-Planck regime (where $s<l_{min}$), it was  argued that the space-time dissolves completely, leading 
the spectral dimension to be zero.

In \cite{nc3}, change of spectral dimension of $\kappa$-Minkowski space-time was studied.  Staring from the Euclidean 
momentum space associated with the $\kappa$-Minkowski space-time, possible Laplacians in the momentum space
were constructed. These Laplacians were constructed by demanding them to be the Casimirs of the 
$\kappa$-Poincare algebra in the momentum space. These Laplacians were constructed as the Casimirs using 
bi-covariant differential calculus and bi-crossproduct basis, respectively. A third form for the Lpalacian 
was derived as the geodesic distance in the $\kappa$-momentum space. By calculating the return probability, 
using these Laplacians in the momentum space, spectral dimensions for these three cases were calculated. 
For the first case, i.e., for the Laplacian written in bi-covariant differential calculus, it was shown 
that the spectral dimension flow from $4$ at low energies to $3$ in the high energies. For the second situation 
where the Laplacian was written in bi-crossproduct basis,  spectral dimension was shown to increase from $4$ to 
$6$ with energy. In the third case, Laplacian was constructed using the notion of relative locality, and it was shown that the 
spectral dimension goes to infinity ($\infty$) as energy increases. Thus 
in the first case, the spectral dimension flows to a lower value (of $3$), replicating the general feature of 
the dimensional flow in other studies, whereas, in the second case, spectral dimension increase to a higher, but 
finite value indicating super diffusion. In the third situation, notion of diffusion breaks down completely 
at high energies as the spectral dimension goes to infinity. A similar behaviour was also noted in \cite{magc} 
for certain  effective models involving non-local laplacians. In the first two cases considered in \cite{nc3}, 
the behaviour of spectral flows were argued to be due to the modified integration measure, necessitated 
by the $\kappa$-deformation. In the third case studied in \cite{nc3} where the spectral dimension is calculated for a 
model compatible with relative locality, the effects of this novel notion of locality leads to higher powers of momenta 
in the dispersion relation. By numerical methods, spectral dimension was calculated and shown that it diverges as 
the probe scale goes to zero.

We note that the dimensional flow in the non-commutative space-time, and in particular for the case of 
$\kappa$-deformed space-time depends on the choice of Laplacians. Also, all these studies\cite{nc1,nc2,nc3,dsr}
where the variation of spectral dimension of $\kappa$-space-time was analysed, used the Laplacians in the momentum 
space. Also, the invariant measure in the momentum space had an important role in deciding the behaviour of spectral
dimension as a function of the probe scale\cite{nc1,nc3}. In \cite{dm,asg,loop}, solutions to diffusion equation were
obtained in the coordinate space and thus it is of interest to analyse the change in the spectral dimension of 
$\kappa$-space-time also using the probability density of the test particle undergoing diffusion in coordinate space. We 
take up this issue here.

In this paper, we derive the spectral dimension of the $\kappa$-space-time, by setting up the $\kappa$-deformed
diffusion equation in terms of commutative co-ordinates. This is achieved by first writing down the $\kappa$-deformed 
Laplacian in the Euclidean version of the $n$-dimensional $\kappa$-space-time, in terms of the derivatives with 
respect to commutative coordinates. We then obtain the solution to this heat equation, perturbatively. Using this 
solution, we calculate the return probability, valid upto second order in the deformation parameter. We then calculate 
the spectral dimension, as a function of diffusion length and the deformation parameter $a$. We then analyse the 
variation of the spectral dimension as a function of diffusion length. We have analysed this for another possible 
realisation of the $\kappa$-deformed Laplacian operator also. Since the framework we use here allow us to set up 
the diffusion equation valid for the $\kappa$-deformed space-time, entirely in the commutative space-time, we can 
use the well established methods to solve this deformed heat equation. Also, we can use the same initial condition 
of demanding the test particle to be  localised at a fixed point in space at the initial time, as in the commutative 
space-time. We have also investigated the effect of finite extension of the probe on the spectral 
dimension, by using a modified initial condition.

This paper is organised as follows. In the next section, a brief summary of 
the $\kappa$-deformed Laplacian operators written in commutative space-time is given\cite{sm}. In section 3, we 
set up the heat equation for a fiducious   particle in this $\kappa$-deformed space-time and obtain its solution. 
This solution is obtained as a perturbative series in the deformation parameter and we obtain the solution valid 
upto first non-vanishing terms in the deformation parameter. Using this solution, we calculate the return probability 
and spectral dimension which is valid upto second order in the deformation parameter. We also discuss various limits 
of the spectral dimension. Using a different initial condition, we calculate the spectral dimension with a probe having 
finite extension. We find that the genric feature of the spectral dimension is not altered by the extended nature of 
the probe. In section 4, we start with a different, possible Laplacian in the $\kappa$-deformed space-time and 
evaluate corresponding spectral dimension and discuss its limits. Here again, we investigate the effect of a probe with 
a finite extension on the spectral dimension. Our concluding remarks are given in section 5.

\section{Kappa-deformed Laplacian in Euclidean space-time}

The approach we take here is to express the $\kappa$-deformed Laplacian in terms of the derivatives with respect to the 
commutative coordinates and deformation parameter, developed in\cite{sm, sm1}.  In this framework, one first re-express the 
coordinates ${\hat x}_\mu$ of the $\kappa$-deformed space-time in terms of the commutative coordinates $x_\mu$ and 
their derivatives. By demanding that this mapping preserves the defining relations satisfied by the non-commutative 
coordinates given by
\begin{equation}
  [\hat{x}_\mu ,\hat{x}_\nu]=i(a_\mu \hat{x}_\nu-a_\nu \hat{x}_\mu)
\end{equation}
where $a_\mu$ are constant real parameters, which describe the deformation of Euclidean space. 
We choose $a_i=0$, i=1 to n-1, and $a_n=a$. Then the algebra of the non-commutative coordinates becomes 
\begin{equation}
 [\hat{x}_i , \hat{x}_j] =0 ,~~~~~~~  [\hat{x}_n , \hat{x}_i] = ia \hat{x}_i,~~~~~~~i,j=1,2,...,n-1
\end{equation}
The explicit form of ${\hat x}_n$ and ${\hat x}_i$ are
\begin{eqnarray}
 {\hat x}_n&=&x_n\psi(A)+iax_i\partial_i\gamma(A),\\
 {\hat x}_i&=&x_i\varphi(A),
\end{eqnarray}
where $A=ia\partial_n$. By demanding that the deformed Poincare transformations must be linear in the coordinates 
and their derivatives, one obtain the generators of the underlying symmetry algebra. These modified generators 
satisfy the same relations as the generators of the usual Poincare algebra and known as undeformed $\kappa$-Poincare 
algebra. These deformed generators are given in terms of the commutative coordinates $x_\mu$ and their derivatives 
$\partial_\mu$ and depend on the deformation parameter $a$. In the limit of vanishing $a$, one recover the Poincare algebra.

The derivatives which transform as vector under $\kappa$-Poincare algebra, called Dirac derivatives, are  
explicitly 
\begin{equation}
  D_i=\partial_i \frac{e^{-A}}{\varphi},~~~~~ D_n=\partial_n\frac{sinhA}{A}+ia\nabla^2 \frac{e^{-A}}{2\varphi^2}.
\end{equation}
The quadratic Casimir of this algebra is $D_\mu D^\mu$ \cite{sm,sm1} is given by 
\begin{equation}
   D_\mu D_\mu = D_i D_i + D_n D_n =\square (1-\frac{a^2}{4} \square ),\label{klaplacian}
\end{equation}
where
\begin{equation}
  \square = \nabla^2 \frac{e^{-A}}{\varphi^2}-\partial_{n}^{2}\frac{2(1-coshA)}{A^2}.\label{box}
\end{equation}
Here $A=ia\partial_n$ and without lose of generality, we choose $\varphi = e^{-\frac{A}{2}}$. Note that the 
$\square$-operator is quadratic in space derivatives and thus $D_\mu D_\mu$ has quartic space derivatives.

Generalising the notion of Laplacian being the Casimir of the Poincare algebra to the $\kappa$-deformed space-time, we
use $D_\mu D_\mu$ as the $\kappa$-deformed Laplacian. In the commutative limit ($a\to 0$), we recover the standard 
Laplacian in the commutative space-time. But if we relax this condition and require only that the $\kappa$-deformed 
operator should reduce to the usual Laplacian in the commutative space-time, we can use the $\square$-operator defined
in eqn.(\ref{box}) also as the $\kappa$-deformed Laplacian. Both $D_\mu D_\mu$ and $\square$ operators have been 
analysed as possible generalisations of Klein-Gordon operator in $\kappa$-space-time\cite{trg1,trg2}.

It is clear from eqn.(\ref{box}) that the $\square$-operator has higher derivatives in time. Further, we note that this 
operator also has terms involving products of derivatives with respect to time and space coordinates. Since 
$D_\mu D_\mu$ is expressed in terms of the $\square$-operator, these properties are carried over to 
$D_\mu D_\mu$ also. This will have important consequences in the calculation of spectral dimension of the 
$\kappa$-deformed space-time. 

Note that both the operators $D_\mu D_\mu$ and $\square$ are written completely in the commutative space-time and thus
the Laplacians we use are expressed in the commutative space-time.

\section{Spectral dimension from $D_\mu D_\mu$ operator}

In this section, we calculate and analyse the spectral dimension of the $\kappa$-deformed space-time, using the 
Casimir of the undeformed $\kappa$-Poincare algebra $D_\mu D_\mu$ as the Laplacian operator, in the $n$-dimensional
$\kappa$-deformed Euclidean space. For this, we start with the diffusion equation in the $\kappa$-deformed space and 
derive its solution. We solve this deformed diffusion equation, perturbatively and obtain the solution valid upto 
second order in the deformation parameter. Using this deformed probability density, we calculate the return probability
and then spectral dimension.

We start with the diffusion equation in an n-dimensional, $\kappa$-deformed Euclidean space, given by
\begin{equation}
    \frac{\partial}{\partial \sigma}U(x,y;\sigma) = D_{\mu} D_{\mu} U(x,y;\sigma).\label{ncde1a}
  \end{equation}
We restrict our attention to first non-vanishing corrections due to non-commutativity. Thus we re-write the above 
equation, valid upto the second order in the deformation parameter $a$ as
  \begin{equation}
    \frac{\partial U}{\partial \sigma} = \nabla^2 U + \partial_n ^2 U -\frac{a^2}{3} \partial_n ^4 U -\frac{a^2}{2} 
    \nabla^2 \partial_n ^2 U - \frac{a^2}{4} \nabla ^4 U \label{NCDE}.
  \end{equation}
Note that all the $a$ dependent terms are of higher derivatives; two of them having quartic derivatives while another
 involves product of quadratic derivatives in space and time.   
  
For convenience we define the Laplacian in the $n$-dimensional commutative Euclidean space-time as
  \begin{equation}
    \nabla^2 U + \partial_n ^2 U = \nabla _n ^2 U.
  \end{equation}
Using this, eqn.(\ref{NCDE}) is re-written as
  \begin{equation}
    \frac{\partial U}{\partial \sigma} = \nabla_n ^2 U  -\frac{a^2}{3} \partial_n ^4 U -\frac{a^2}{2}  \nabla^2  
    \partial_n ^2 U - \frac{a^2}{4} \nabla ^4 U \label{MDE}.
  \end{equation}
   We use perturbative approach to solve this equation to obtain  the heat kernel U(x,y;$\sigma$). Thus we start with 
 the  probability density, valid upto second order in the deformation parameter $a$ as
   \begin{equation}
    U=U_0+aU_1+a^2 U_2.\label{PS}
  \end{equation}
Note that we have the following relations between dimensions of terms in the above perturbative series,
  \begin{equation}
    [U_1]=\frac{1}{L}[U_0]
  \end{equation}
  and
  \begin{equation}
    [U_2]=\frac{1}{L^2}[U_0].
  \end{equation}   
Using eqn.(\ref{PS}) in eqn.(\ref{MDE}) and equating the zeroth order terms in $a$, we find that $U_0$ satisfy 
the usual heat equation,
  \begin{equation}
    \frac{\partial}{\partial \sigma} U_0(x,y;\sigma)=\nabla^2_n U_0(x,y;\sigma).\label{u0eqn}
  \end{equation}
The solution to above equation is
  \begin{equation}
    U_0(x,y;\sigma)=\frac{1}{(4\pi \sigma)^\frac{n}{2}} e^{-\frac{\mid x-y \mid^2 }{4\sigma}}\label{U_0}.
  \end{equation}
Next, we equate the first order terms in `a' in eqn.(\ref{MDE}) to obtain
  \begin{equation}
    \frac{\partial}{\partial \sigma} U_1(x,y;\sigma)=\nabla^2_n U_1(x,y;\sigma)\label{u1eqn}
  \end{equation}
  showing that $U_1$ also satisfy the same heat equation as $U_0$. Thus we find $U_1$ to be of the same form as
  $U_0$, i.e., 
  \begin{equation}
    U_1(x,y;\sigma)=\frac{\alpha}{(4\pi \sigma)^\frac{n}{2}} e^{-\frac{\mid x-y \mid^2 }{4\sigma}}\label{U_1}
  \end{equation}
where $\alpha$ has the dimensions of $L^{-1}$. Thus, for solving both $U_0$ and $U_1$, we have used the usual initial condition, i.e.,
$U_0(x,y;0)=\delta^n(x-y)=U_1(x,y;0)$. This is possible because we could write the deformed diffusion equation in the
commutative space-time.

Now we solve for the next term in eqn.(\ref{PS}). For this we equate the second order terms in `a' 
in eqn.(\ref{MDE}) and obtain
  \begin{equation}
    \frac{\partial}{\partial\sigma}U_2(x,y;\sigma)=\nabla_n ^2 U_2(x,y;\sigma)-\frac{1}{3}\partial_n ^4 U_0(x,y; \sigma) -\frac{1}{2} \nabla^2 \partial_n ^2 U_0(x,y;\sigma)-\frac{1}{4} \nabla ^4 U_0(x,y;\sigma) \label{u2eqn}
  \end{equation}
  Note that the last three terms on the RHS of above equation show the change in the diffusion equation due to the
  $\kappa$-deformation. Using the solution for $U_0$ obtained in eqn.(\ref{U_0}) in the above ,
and after straight forward simplification, we get the equation satisfied by $U_2$ as
  \begin{equation}
   \begin{split}
    \frac{\partial}{\partial \sigma}U_2(x,y;\sigma) = \nabla_n ^2 U_2(x,y;\sigma) + 
    [-\frac{(n+1)^2}{16\sigma^2} + \frac{(x_n-y_n)^2}{16 \sigma^3} + \frac{n+2}{16 \sigma^3}\Sigma _{i=1}^{n}(x_i-y_i)^2 - 
    \frac{(x_n -y_n)^4}{48\sigma^4}\\-\frac{(x_n-y_n) ^2}{32 \sigma^4}\Sigma _{i=1}^{n-1}(x_i-y_i)^2 -\frac{1}{64 \sigma^4}
    (\Sigma _{i=1}^{n-1}(x_i-y_i)^2)^2] \frac{1}{(4\pi\sigma)^{\frac{n}{2}}}e^{-\frac{\mid x-y \mid^2 }{4\sigma}}.
   \end{split}
  \end{equation}
Using  Duhamel's principle\cite{dp}, we solve the above equation which is of the generic form
  \begin{equation}
    \frac{\partial}{\partial \sigma}U_2(X,\sigma)=\nabla^2_n U_2(X,\sigma)+f(X,\sigma)\label{dm}
  \end{equation}
  where $ X = x-y $. With the initial condition
  \begin{equation}
    U_2(X,0)=g(X),
  \end{equation}
  the solution to eqn. (\ref{dm}) is given by
  \begin{equation}
    U_2(X,\sigma)=\int_{R^n} \Phi (X-X',\sigma)g(X')dX' + 
    \int_{0}^{\sigma}\int_{R^n} \Phi (X-X',\sigma-s) f(X',s) dX' ds \label{U_2}
  \end{equation}
where 
  \begin{equation}
    \Phi(X,\sigma)=\frac{1}{(4 \pi \sigma)^{\frac{n}{2}}} e^{-\frac{\mid X \mid^2}{4\sigma}}\label{phi}.
  \end{equation}
In our case we have the initial condition satisfied by the solution as
\begin{equation}
 U_2(X,0)=g(X) = \delta^n(X).\label{bc}
 \end{equation}
Note that we are using the same boundary condition as in the usual diffusion equation. This is to be contrasted with
the approach taken in \cite{nc2}.  Since, the $\kappa$-deformed Laplacian and hence the diffusion equation
are written fully in the commutative space-time and all the effects of the non-commutativity is
included in the modified Laplacian and thus we are justified in using this initial condition.

Using eqn(\ref{phi}) and eqn.(\ref{bc}), the first term on RHS of eqn.(\ref{U_2}), $U_{21}$ is calculated as 
\begin{equation}
 U_{21}(x,y;\sigma) = \int_{R^n} \Phi (X-X',\sigma)g(X')dX' = 
 \frac{\beta}{(4 \pi \sigma)^{\frac{n}{2}}} e^{-\frac{\mid x-y \mid^2}{4 \sigma}}\label{U_21},
\end{equation}
where $\beta$ has the dimensions of $L^{-2}$. The second term on RHS of eqn.(\ref{U_2}), $U_{22}$ is evaluated as
  \begin{eqnarray}
U_{22}(x,y,\sigma) &=& \int_{0}^{\sigma}\int_{R^n} \Phi (X-X',\sigma-s) f(X',s) dX' ds \nonumber\\
&  =& \frac{1}{(4 \pi \sigma)^\frac{n}{2}}e^{- \frac{\mid x-y \mid ^2}{4 \sigma}}
   [\left(-\frac{(x_n-y_n)^4}{48 \sigma^4}-\frac{(x_n-y_n)^2}{32 \sigma^4}\Sigma _{i=1}^{n-1}(x_i-y_i)^2-
  \frac{1}{64 \sigma^4}\left(\Sigma _{i=1}^{n-1}(x_i-y_i)^2\right)^2\right)(\sigma-\epsilon) \nonumber\\ 
&  +& \left(\frac{(x_n-y_n)^2}{16 \sigma^2}+\frac{n+2}{16 \sigma^2}\Sigma _{i=1}^{n}(x_i-y_i)^2\right)
  \ln (\sigma/\epsilon)+ \left(\frac{(n+1)^2}{16}\right)\left(\frac{1}{\sigma}-\frac{1}{\epsilon }\right)\nonumber\\
  & -&\left( \frac{(x_n-y_n)^2}{4\sigma^3} +\frac{1}{16 \sigma^3} \Sigma _{i=1}^{n-1}(x_i-y_i)^2 
+\frac{(n-1)}{16 \sigma^3}(x_n-y_n)^2 + \frac{(n+1)}{16 \sigma^3} \Sigma _{i=1}^{n-1}(x_i-y_i)^2 \right){\mathds{A}}
  \nonumber\\
  &  -& \frac{1}{2 \pi \sigma^3} \left((x_n-y_n) \Sigma _{i=1}^{n-1}(x_i-y_i)+(x_1-y_1)\Sigma_{i=2}^{n-1}(x_i-y_i)+...
  +(x_{n-2}-y_{n-2})(x_{n-1}-y_{n-1})\right){\mathds {A}}
  \nonumber\\
&  -&
  \left(\frac{(x_n-y_n)^3}{6 \sigma^3 \sqrt{\sigma \pi}}+\frac{(x_n-y_n)}{8 \sigma^3 \sqrt{\sigma \pi}}
  \Sigma _{i=1} ^{n-1} (x_i-y_i) ^2 + \frac{(x_n-y_n)^2}{8 \sigma^3 \sqrt{\sigma \pi}} 
  \Sigma _{i=1} ^{n-1} (x_i-y_i) \right){\mathds{B}}\nonumber \\
  & -& \left(\frac{1}{8 \sigma ^3 \sqrt{\sigma \pi}}\Sigma_{i=1}^{n-1}(x_i-y_i)
  \Sigma_{j=1}^{n-1} (x_j-y_j)^2\right){\mathds{B}}\nonumber \\
  &  + &\left(\frac{2 (x_n-y_n)}{3 \sigma^2 \sqrt{\sigma \pi}} - \frac{(n+1)}{4 \sigma^2 \sqrt{\sigma \pi}} 
  \Sigma_{i=1}^{n-1}(x_i-y_i) -\frac{(n-1)}{4 \sigma^2 \sqrt{\sigma \pi}}(x_n-y_n)\right)
    \left((2 \sigma+\epsilon) 
  \sqrt{\frac{\sigma}{\epsilon }-1}-3 \sigma \tan^{-1}\sqrt{\frac{\sigma}{\epsilon }-1}\right) 
  \nonumber\\
&  -&
  \frac{(n+1)^2}{16\sigma^2}\left(-2\sigma \ln (\sigma/\epsilon) +\frac{\sigma^2}{\epsilon }-\epsilon \right)
  +
  \frac{(n+1)^2}{8 \sigma} \left[-1 + \frac{\sigma}{\epsilon }
  - \ln(\sigma/\epsilon)\right] \nonumber\\
   &+&2\left(\frac{(x_n-y_n)}{4\sigma\sqrt{\sigma \pi}}+ 
  \frac{n+2}{4 \sigma \sqrt{\sigma \pi}}\Sigma _{i=1}^{n}(x_i-y_i)\right)\left(tan^{-1}(\sqrt{\sigma/\epsilon-1})-
  \sqrt{\sigma/\epsilon-1}\right)]\label{U_22}
  \end{eqnarray}
 where ${\mathds {A}}=\sigma \ln (\sigma/\epsilon) -\sigma +\epsilon $ and 
 ${\mathds{B}}=(\sigma \tan^{-1}\sqrt{\frac{\sigma}{\epsilon }-1} - 
  \epsilon \sqrt{\frac{\sigma}{\epsilon }-1})$.
 
 Note that $U_{22}$ treats Euclidean time and space coordinate on a different footing. This is apparent from the fact 
 that we have terms depending on $(x_n-y_n)$ and $(x_i-y_i)$, separately. The $\epsilon$ appearing in the above is a 
 lower cut-off introduced in evaluating the integral in eqn.(\ref{U_2}) and we will set the limit $\epsilon\to 0$ after calculating 
 the spectral dimension.
  
Using eqn.(\ref{U_21}) and eqn.(\ref{U_22}), we get the solution to the second order correction as
$ U_2(x,y;\sigma) = U_{21}(x,y;\sigma) +U_{22}(x,y;\sigma)  $. Using eqns. (\ref{U_0}), (\ref{U_1}), (\ref{U_21})
and eqn.(\ref{U_22}) in eqn.(\ref{PS}),  we find heat kernel, valid upto second order in a.  

Using this in eqn.(\ref{rp}), we obtain the return probability as
\begin{equation}
 P(\sigma)= \frac{1}{(4 \pi \sigma)^{\frac{n}{2}}}\left[1+ a \alpha+ a^2 \beta 
 +a^2 \left(-\frac{(n+1)^2}{16 \sigma}+\frac{(n+1)^2}{16 \sigma^2}\epsilon \right)\right].
\end{equation}
Note that the return probability, calculated above, do have first order as well as second order corrections due to the 
non-commutativity. 

Using this in eqn.(\ref{specdim}), we evaluate the spectral dimension as
\begin{equation}
  D_s = \frac{n+na\alpha+na^2 \beta-(n+2)(n+1)^2 \frac{a^2}{16\sigma}+(n+4)(n+1)^2
  \frac{a^2}{16 \sigma^2}\epsilon }{1+a\alpha+ a^2 \beta- a^2 \frac{(n+1)^2}{16\sigma}(1-\frac{\epsilon }{\sigma})}.
\end{equation} 
Keeping upto first non-vanishing terms, we obtain the spectral dimension as $ D_s = n-
(n+1)^2\frac{a^2}{8\sigma}+(n+1)^2\frac{a^2}{4\sigma^2}\epsilon.$ After setting the cut-off parameter $\epsilon$ to zero, 
we obtain spectral dimension of the $\kappa$-deformed space-time as
\begin{equation}
  D_s = n-(n+1)^2\frac{a^2}{8\sigma}.\label{specdim1}
\end{equation}
Note that the first non-vanishing correction due to non-commutativity is in the second order in $a$ and this correction 
also depends on the initial topological dimension $n$(apart from the diffusion parameter $\sigma$).

\begin{figure}
\caption{ spectral dimension as a function of $\sigma$ with $a=1$ and $n=4$ for $D_\mu D_\mu$ as Laplacian.}\label{fig1}
 \includegraphics[height=2 in, width=2in]{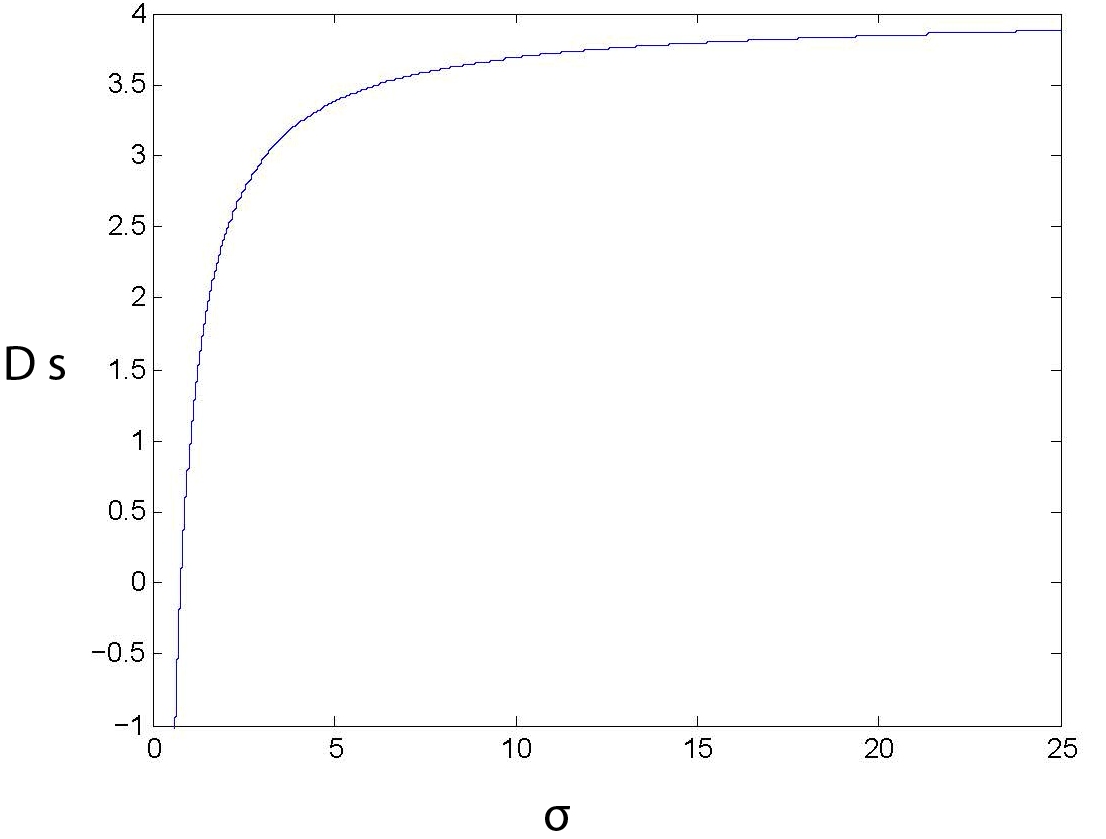}
\end{figure}

We see that in the limit of large diffusion parameter, the spectral dimension(see. Fig.\ref{fig1}) is same as the topological dimension, i.e.,
\begin{equation}
 Lt_{\sigma\to\infty}D_s=n.
\end{equation}
In the limit of vanishing $\sigma$, we find that the spectral dimension goes to $-\infty$. For $a=1$ and $n=4$,
we find the spectral dimension to be zero for $\sigma=0.78$. In general, for $n=4$, the spectral dimension becomes 
negative for $\sigma<25a^2/32$. As in \cite{nc2}, we see here that for trans-Planckian regime where 
$\sigma=25a^2/32$, space-time 
dissolves, setting the spectral dimension to be zero. Since the deformation parameter $a$ is related to the minimum 
length scale associated with the $\kappa$-space-time, negative spectral dimension for $\sigma<25a^2/32$ seems to 
indicate that the diffusion equation and/or spectral dimension looses its meaning below this scale of $25a^2/32$.

It is well known that the Laplacians with higher derivatives do lead to negative probability\cite{gc1} and here, as 
we noted, our Laplacian do have higher derivatives in space as well as time coordinates. In the spirit of \cite{gc}, we can take 
the view that the $\kappa$-deformed diffusion equation  eqn.(\ref{NCDE}) and hence the spectral dimension obtained 
in eqn.(\ref{specdim1}) is valid only for $a^2<32\sigma/25$ and thus assure that the spectral dimension is 
positive definite. Since the deformation parameter $a$ is expected to be of the order of Planck length, probing well 
below the scale of the order of $a^2$ may not be possible and the existence of such a threshold for the probe scale is 
plausible.

%%%%%%%%%%%%%%%%%%%%%%%%%%%%%%%%%%%%%%%%%%%%%%%%%%%
In \cite{nc2}, spectral dimension for a non-commutative space-time was studied by using  an extended probe and analysed the diffusion equation satisfied by this probe and spectral dimension was calculated.
In our case, we have started with the diffusion equation in the $\kappa$-deformed space-time, written in commutative space-time and  effects of non-commutativity was incorporated through higher derivative terms appearing
in the $\kappa$-deformed Laplacian(see eqn.(\ref{NCDE})).  Here, since the diffusion equation is written in the commutative space-time itself, we have used point particle as the probe. We can also use an extended 
probe in this approach also. The extended nature of the probe can be incorporated using an initial condition which takes into account of this finite length of the probe particle.  For this, we first solve
for  $U_0$ satisfying eqn.(\ref{u0eqn}) with a modified initial condition given by
   \begin{equation}
    U_0(x,y;0)=\frac{1}{(4\pi a^2)^\frac{n}{2}} e^{-\frac{\mid x-y \mid^2 }{4 a^2}}.\label{INITIAL}
  \end{equation}
  Note that the width of the Gaussian above depends on the deformation parameter $a$ which has the dimension of length.  With this initial condition, we get 
  \begin{equation}
    U_0(x,y;\sigma)=\frac{1}{(4\pi (\sigma+a^2))^\frac{n}{2}} e^{-\frac{\mid x-y \mid^2 }{4(\sigma+a^2)}}.
  \end{equation}
Since we are interested in the solution valid upto second order in $a$, keeping terms  only up to second order terms in `$a$'  we find
 \begin{equation}
   U_0(x,y;\sigma)=\frac{1}{(4\pi \sigma)^\frac{n}{2}} e^{-\frac{\mid x-y \mid^2 }{4\sigma}}\left[1-\frac{n a^2}{2\sigma}+\frac{a^2 \mid x-y\mid^2}{4\sigma^2}\right] .\label{NEWU_0}
 \end{equation}
 Next, we solve eqn.(\ref{u1eqn})  using the initial condition given in eqn.(\ref{INITIAL})  and obtain   
  \begin{equation}
    U_1(x,y;\sigma)=\frac{\alpha}{(4\pi (\sigma+a^2))^\frac{n}{2}} e^{-\frac{\mid x-y \mid^2 }{4(\sigma+a^2)}}
  \end{equation}
Here  we keep terms up to first order in $a$ since the solution $U$ contains $aU_1$. Thus $U_1$, valid up to first order in $a$ is given by
 \begin{equation}
    U_1(x,y;\sigma)=\frac{\alpha}{(4\pi \sigma)^\frac{n}{2}} e^{-\frac{\mid x-y \mid^2 }{4\sigma}} .
  \end{equation}
Next we solve for $U_2$ satisfying eqn.(\ref{u2eqn}). In order to solve this we use $U_0$ obtained in eqn.(\ref{NEWU_0}). We also use the initial condition give in eqn.(\ref{INITIAL}), which incorporate
 the extended nature of the probe. But  in this case,  we  need to consider only zeroth order terms in `$a$' (since $U_2$ comes with a coefficient $a^2$ in $U$) and thus the solution for $U_2$, valid up 
 to zeroth order in $a$, is same as the one satisfying eqn.(\ref{u2eqn}), obtained earlier.

Thus, with the probe having an extension,  we find the solution for heat kernel, valid upto second order in `$a$' as
\begin{equation}
  U(x,y;\sigma)=\frac{1}{(4\pi \sigma)^\frac{n}{2}} e^{-\frac{\mid x-y \mid^2 }{4\sigma}}\left[1-\frac{n a^2}{2\sigma}+\frac{a^2 \mid x-y\mid^2}{4\sigma^2}+a \alpha+a^2 \beta\right]+a^2U_{22}.
\end{equation}
Using this solution we obtain the return probability as
\begin{equation}
 P(\sigma)= \frac{1}{(4 \pi \sigma)^{\frac{n}{2}}}\left[1+ a \alpha+ a^2 \beta 
 -a^2 \frac{(n+1)^2}{16 \sigma}-\frac{n a^2}{2 \sigma}\right].
\end{equation}
Using this, we evaluate the spectral dimension to be
\begin{equation}
  D_s = \frac{n+na\alpha+na^2 \beta-(n+2)(n+1)^2 \frac{a^2}{16\sigma}-n(n+2)
  \frac{a^2}{2 \sigma}}{1+a\alpha+ a^2 \beta- a^2 \frac{(n+1)^2}{16\sigma}-\frac{n a^2 }{2 \sigma})}.
\end{equation}
Keeping upto first non-vanishing terms in `$a$', we find 
\begin{equation}
  D_s = n-(n+1)^2\frac{a^2}{8\sigma}-\frac{n a^2}{\sigma} \label{specdim1a}
\end{equation}

By comparing with the eqn.(\ref{specdim1}) we see that the extended probe leads to  an additional term $-\frac{na^2}{\sigma}$. Note that this additional term depends on the topological dimension $n$. 
We note that, as for the particle like probe used earlier, here too,  the spectral dimension is same as the topological dimension in the limit $\sigma \rightarrow \infty$. Similarly, in the limit
 $\sigma \rightarrow 0$ we  see that $D_s \rightarrow -\infty$, exactly same as in the case of particle like probe used 
 in obtaining spectral dimension obtained in eqn.(\ref{specdim1}).
 
For $a = 1$ and $n = 4$, we find the spectral dimension obtained in eqn.(\ref{specdim1a}) vanishes when $\sigma = 1.781$. For $n = 4$, the spectral dimension become negative for $\sigma < \frac{57 a^2}{32}$.  In general, the spectral dimension vanishes for $\sigma = \frac{57 a^2}{32}$.  By demanding that the spectral dimension should be positive definite imply  a cut-off on the deformation parameter given by $a^2 < \frac{32 \sigma}{57}$.  Thus we see that the effect of extended 
nature of the probe modify the cut-off values of $\sigma$ and $a^2$, but do not change the general feature of the dimensional flow and also do not 
affect the values of spectral dimension in the limit  $\sigma \rightarrow 0$ as well as in the limit $\sigma \rightarrow \infty$.

\section{Spectral dimension from $\square$ operator}

In this section, we use $\square$ operator as the Laplacian in the $\kappa$-deformed space-time, which reduces to the usual Laplacian in the commutative space-time. Thus we start with the diffusion equation 
  \begin{equation}
    \frac{\partial}{\partial \sigma}U(x,y;\sigma) = \square U(x,y;\sigma),\label{ncde2a}
  \end{equation}
where $ \square = \nabla^2 \frac{e^{-A}}{\varphi^2}-\partial_{n}^{2}\frac{2(1-coshA)}{A^2}$,
and $A=ia\partial_n $. Also, we choose $\varphi=e^{-\frac{A}{2}}$. Keeping upto first non-vanishing terms in $a$, 
the above diffusion equation becomes
\begin{equation}
\frac{\partial}{\partial \sigma}U(x,y;\sigma)=\left(\nabla^{2}_{n}-\frac{a^2}{12}\partial_n^4\right)U(x,y;\sigma).
\label{ncde2}
\end{equation} 
Here, we note that the $a$ dependent term involves quartic time derivative, but there are no terms involving
product of derivatives with respect to different coordinates. As earlier, we solve the above equation 
perturbatively using eqn.(\ref{PS}) for $U$. Collecting the zeroth order terms in $a$ shows that $U_{0}$ satisfy 
the usual heat equation, i.e.,
\begin{equation}
    \frac{\partial}{\partial \sigma} U_0(x,y;\sigma)=\nabla^2_n U_0(x,y;\sigma)\label{u0eq}
  \end{equation}
with solution
  \begin{equation}
    U_0(x,y;\sigma)=\frac{1}{(4\pi \sigma)^\frac{n}{2}} e^{-\frac{\mid x-y \mid^2 }{4\sigma}}\label{BU_0}
  \end{equation}
and the first order terms in $a$ leads to the equation 
  \begin{equation}
    \frac{\partial}{\partial \sigma} U_1(x,y;\sigma)=\nabla^2_n U_1(x,y;\sigma)\label{u1eq}
  \end{equation}
showing that $U_{1}$ also satisfy the usual heat equation and therefore the solution is 
  \begin{equation}
    U_1(x,y;\sigma)=\frac{\alpha}{(4\pi \sigma)^\frac{n}{2}} e^{-\frac{\mid x-y \mid^2 }{4\sigma}}
  \end{equation}
where $[\alpha]$ has the dimensions of $L^{-1}$. Note that in obtaining the solutions for $U_0$ and $U_1$, we 
have used the initial condition of the test particle being localised at a given point in the space, as in the
commutative space-time.

Next, by equating the second order terms in $a$, we find that the equation satisfied by $U_{2}$ as
\begin{equation}
\frac{\partial}{\partial \sigma}U_2(x,y;\sigma)=\nabla^2_n U_2(x,y;\sigma)-\frac{1}{12}\partial_n^4 U_0(x,y;\sigma).\label{u2eq}
\end{equation}
Note that the above equation has a quartic term in the Euclidean time derivative. Now substituting eqn.(\ref{BU_0}) in the above,
we obtain $U_2$ as
\begin{eqnarray}
U_2(x,y;\sigma)&=&\frac{e^{-\frac{|x-y|^2}{4\sigma}}}{(4\pi\sigma)^{\frac{n}{2}}}
[\beta-\frac{(x_n-y_n)^4}{192\sigma^4}(\sigma-\epsilon) +\frac{(x_n-y_n)^2}{16\sigma^2}\ln(\sigma/\epsilon)
+\frac{(\epsilon-\sigma)}{16\sigma\epsilon}\nonumber\\
&-&\frac{(x_n-y_n)^2}{16\sigma^3}\left(\sigma\ln(\sigma/\epsilon)-\sigma+\epsilon\right)-
\frac{(x_n-y_n)^3}{24\sqrt{\sigma^7\pi}}
\left(\sigma tan^{-1}\sqrt{\sigma/\epsilon-1}-\epsilon\sqrt{\sigma/\epsilon-1}\right)\nonumber\\
&-&\frac{(x_n-y_n)}{6\sqrt{\sigma^5\pi}}\left[(2\sigma+\epsilon)\sqrt{\sigma/\epsilon-1}
-3\sigma tan^{-1}{\sqrt{\sigma/\epsilon-1}}\right]-\frac{1}{16\sigma^2}\left(-2\sigma\ln(\sigma/\epsilon)
+\frac{\sigma^2-\epsilon^2}{\epsilon}\right)\nonumber\\
&+&\frac{1}{8\sigma}\left[-1-\ln(\sigma/\epsilon)+\sigma/\epsilon\right] +
\frac{(x_n-y_n)}{2\sqrt{\sigma^3\pi}}\left(tan^{-1}\sqrt{\sigma/\epsilon-1}-\sqrt{\sigma/\epsilon-1}\right)].
 \end{eqnarray}
As in eqn.(\ref{U_22}), we see that the Euclidean time and space coordinate are treated differently. Here also, 
the $\epsilon$ is the lower cut-off introduced in evaluating the integral and we will take the limit 
$\epsilon\to 0$ after calculating the spectral dimension.

Using this in eqn.(\ref{rp}), we obtain the return probability 
\begin{equation}
  P(\sigma)= \frac{1}{(4 \pi \sigma)^{\frac{n}{2}}}\left[1+ a \alpha+ a^2 \beta +a^2 \left(-\frac{1}{16 \sigma}+\frac{\epsilon}{16 \sigma^2}\right)\right]
\end{equation}
and from eqn.(\ref{specdim}), we find the spectral dimension as
\begin{equation}
  D_s = \frac{n+na\alpha+na^2 \beta-(n+2)\frac{a^2}{16\sigma}+(n+4)\frac{a^2\epsilon}{16 \sigma^2}}{1+a\alpha+a^2 \beta+ a^2 (\frac{-1}{16\sigma}+\frac{\epsilon}{16\sigma^2})}.
\end{equation} 
\begin{figure}
 \caption{spectral dimension as a function of $\sigma$ with $a=1$ and $n=4$ for $\square$ as Laplacian.}\label{fig2}
 \includegraphics[height=2in,width=2in]{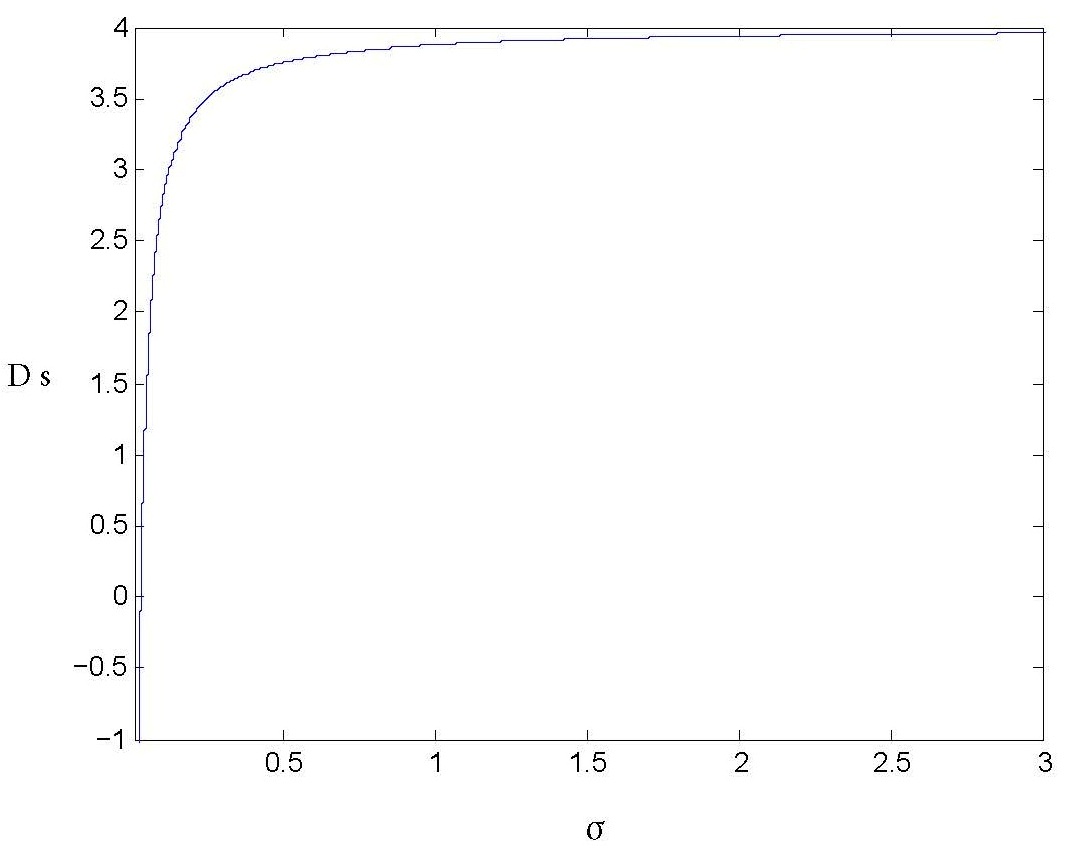}
\end{figure}
We note that all the first order terms in $a$ cancel with each other and thus the first non-vanishing correction due 
to the non-commutativity is in the second order in the deformation parameter $a$. After setting the cut-off parameter 
to zero, we find the spectral dimension to be
\begin{equation}
  D_s = n-\frac{a^2}{8\sigma}.\label{specdim2}
\end{equation}
We note that the non-commutative correction is independent of the topological dimension $n$. Thus the change in the 
effective dimension is same irrespective of the topological dimension of the space-time we start with. This should be
contrasted with the spectral dimension we found in eqn.(\ref{specdim1}).

 It is easy to see(Fig.\ref{fig2}) that in the limit $\sigma\to\infty$, we find that spectral dimension and topological 
dimension to be equal. For $\sigma=a^2/8n$, the spectral dimension vanishes and for smaller values ($\sigma<a^2/8n$), 
the spectral dimension becomes negative. As in the case of Eqn.(\ref{specdim1}), here to, this suggest that the
spectral dimension seems to loose its meaning below this scale. Here again, we note that the deformed Laplacian we used 
here had higher time derivatives and such models are known to lead to negative probability density\cite{gc}. Here again, 
we can set the bound on the deformation parameter as $a^2<8n\sigma$ to guarantee the positivity of the spectral 
dimension.

%%%%%%%%%%%%%%%%%%%%%%%%%%%%%%%%%%%%%%%%%%%%%%

   As in the case  of solving eqn. (\ref{NCDE}), we now solve eqn.(\ref{ncde2}) for extended probe by using a Gaussian function whose width is dictated by the deformation parameter, in place of delta function initial condition used
   earlier.
   
   With the initial condition which imply the extended nature of the probe
   \begin{equation}
    U_0(x,y;0)=\frac{1}{(4\pi a^2)^\frac{n}{2}} e^{-\frac{\mid x-y \mid^2 }{4 a^2}}\label{INITIAL2},
  \end{equation}
  we find  the solution to $U_0$ satisfying  eqn.(\ref{u0eq}), valid up to second order in $a$ as 
  \begin{equation}
   U_0(x,y;\sigma)=\frac{1}{(4\pi \sigma)^\frac{n}{2}} e^{-\frac{\mid x-y \mid^2 }{4\sigma}}\left[1-\frac{n a^2}{2\sigma}+\frac{a^2 \mid x-y\mid^2}{4\sigma^2}\right] .\label{NEW2U_0}
 \end{equation}
Similarly, solving eqn.(\ref{u1eq}) with the  modified initial condition given above, we obtain the solution for $U_1$, valid up to first order in $a$ as
\begin{equation}
    U_1(x,y;\sigma)=\frac{\alpha}{(4\pi \sigma)^\frac{n}{2}} e^{-\frac{\mid x-y \mid^2 }{4\sigma}}. \label{NEW2U_1}
  \end{equation}
  Since we only need solution of $U_2$ valid only up to zeroth order in $a$, we find that the solution is same as the one satisfying eqn.(\ref{u2eq}).
Thus we have the  solution for heat kernel valid upto second order in `$a$' as
\begin{equation}
  U(x,y;\sigma)=\frac{1}{(4\pi \sigma)^\frac{n}{2}} e^{-\frac{\mid x-y \mid^2 }{4\sigma}}\left[1-\frac{n a^2}{2\sigma}+\frac{a^2 \mid x-y\mid^2}{4\sigma^2}+a \alpha+a^2 \beta\right]+a^2U_{22}.
\end{equation}
Using this solution we obtain the return probability as
\begin{equation}
 P(\sigma)= \frac{1}{(4 \pi \sigma)^{\frac{n}{2}}}\left[1+ a \alpha+ a^2 \beta 
 - \frac{a^2}{16 \sigma}-\frac{n a^2}{2 \sigma}\right].
\end{equation}
and the spectral dimension as
\begin{equation}
  D_s = \frac{n+na\alpha+na^2 \beta-(n+2) \frac{a^2}{16\sigma}-n(n+2)
  \frac{a^2}{2 \sigma}}{1+a\alpha+ a^2 \beta- \frac{a^2}{16\sigma}-\frac{n a^2 }{2 \sigma})}.
\end{equation}
Keeping  terms valid  up to first non-vanishing terms in $a$, we find
\begin{equation}
  D_s = n-\frac{a^2}{8\sigma}-\frac{n a^2}{\sigma} 
\end{equation}
By comparing with the eqn.(\ref{specdim2}) we see an additional term $-\frac{na^2}{\sigma}$ in the above, which depends on the topological dimension.  We also note that the limiting values of the spectral dimension is 
same as in that for the spectral dimension obtained for point probe in eqn.(\ref{specdim2}).

For a = 1 and n = 4, we note that the spectral dimension $D_s$  equal to zero when $\sigma = 1.031 $. For  $n = 4$, the spectral dimension become negative for $\sigma < \frac{33 a^2}{32}$. 
Thus, we see that the spectral dimension vanishes for $\sigma = \frac{33 a^2}{32}$.  As earlier the requirement of positivity of the spectral dimension will lead to the  condition $a^2 < \frac{32 \sigma}{33}$. 
 We see that the effect of extended  nature of the probe only modify the cut-off values of $\sigma$ and $a^2$, while the general feature of the dimensional flow and the values of spectral dimension in the limit  
$\sigma \rightarrow 0$ as well as in the limit $\sigma \rightarrow \infty$ are unaffected,  in the case of Laplacian being $\square$ also.

%%%%%%%%%%%%%%%%%%%%%%%%%%%%%%%%%%%%%%%%%%
 
 \section{Conclusion}

In this paper, we have derived the spectral dimension of $\kappa$-deformed space-time and analysed dimensional flow
with the probe scale. Here, we have adopted an approach which allowed us to obtain the diffusion equation valid for the
$\kappa$-space-time, written in terms of derivatives with respect to the commutative coordinates. By using a mapping 
of the coordinates of $\kappa$-deformed space-time to commutative coordinates and their derivatives, the symmetry 
algebra of the $\kappa$-space-time was obtained in \cite{sm,sm1}. The generators of this algebra and hence the Casimir
were explicitly given in the commutative space-time. We have used this Casimir as the $\kappa$-deformed Laplacian 
and this allowed us to set up the deformed diffusion equation which is completely expressed in the commutative 
space-time. Apart from facilitating the use of well established calculational tools, this also allowed us to implement
the initial condition that the test particle is at a fixed point in the space at the initial time ($\sigma=0$). The effect 
of $\kappa$ deformation appeared in the Laplacian through higher derivatives. Keeping terms upto second order in the 
deformation parameter, we solved this equation, perturbatively. After calculating the probability to find the particle 
back at the initial position after the lapse diffusion time $\sigma$, we derived the spectral dimension.  We found 
that the spectral dimension decreases from the topological dimension. For a space-time of dimension $4$, the spectral 
dimension reduces to zero when the probe scale ($\sigma$) is $25 a^2/32$ and for further finer probe scale, spectral dimension 
become negative and goes to $-\infty$ as $\sigma$ vanishes.

We have also calculated the spectral dimension with another choice of $\kappa$-deformed Laplacian. This
choice of Laplacian is dictated only by the requirement of correct commutative limit and it is not a Casimir of the
undeformed $\kappa$-Poincare algebra,  symmetry algebra of the deformed space-time. In this case also, the 
spectral dimension decreases with probe scale and flows to $-\infty$ as $\sigma$ goes to zero. The spectral dimension 
becomes zero when $\sigma=a^2/32$ and for further smaller values it become negative.

 We have studied the changes in the spectral dimension using a probe with a finite extension, for both the choices of the $\kappa$-deformed Laplacian.  In these case also, we find the same behavior for the spectral dimension at the limit of large as well as small probe scale 
 $\sigma$. Thus the generic feature of the dimensional flow is unaffected when the probe has a finite extension.

 The non-commutativity of space-time introduces non-local effects. These effects do appear in the deformed Laplacian. 
In our case these non-local effects appear through the higher derivative terms in the Laplacian used in eqn.(\ref{ncde1a}) and 
eqn.(\ref{ncde2a}). In \cite{dsr,nc1,nc2,nc3} the non-locality appeared through the higher powers of momenta present in the dispersion
relation. Note that the relative locality principle which modifies the notion of locality, also modifies the dispersion
relation used in \cite {nc3} to calculate the spectral dimension.

Vanishing of spectral dimension was argued to be sign of space-time loosing its meaning at 
trans-Planckian scales\cite{nc2}. We  note that the deformation parameter $a$, having length dimension, is related 
to the minimal length associated with the $\kappa$-space-time and expected to be of the order of Planck length. For 
both the cases we studied here, we find that the spectral dimension is positive much below the scale set by the 
deformation parameter $a$ and become negative when the probe scale is less than $0.78a^2$ and $0.031a^2$, respectively.
As discussed in the introduction, some of the earlier studies on the dimensional flow in $\kappa$-space-time, using 
an approach different from the one used here, showed the growth of spectral dimension to $+\infty$, as the probe scale 
vanishes\cite{nc3}, indicating super diffusion at small probe scales. In the present case, we have the opposite 
situation where the spectral dimension goes to $-\infty$.

The probability density becoming negative for diffusion equations involving higher derivatives have been noted earlier
and a possible re-interpretation of the spectral dimension in such scenarios was discussed in \cite{gc}.  It was noted 
that at low energies, the spectral dimension matchs the topological dimension only for certain values of the 
deformation parameter and it was argued that there should be a threshold value of deformation parameter below which the
return probability is not a physically acceptable solution. We can take a similar approach in the cases studied here. 
We have seen that for certain values of the diffusion parameter we get the spectral dimension to be negative. By inverting 
this inequality, we can get condition on the deformation parameter ($a^2<32\sigma/25$ and $a^2<32\sigma$, respectively
for the cases we considered here with $n=4$) for which the spectral dimension is positive. This may also be an indication of the 
existence of multi-scale structure in the $\kappa$-space-time\cite{gc2}. A re-interpretation of the spectral dimension 
was discussed in\cite{gc}, which avoids the negative probability associated  with the higher order, and non-local 
dispersion relations. Since in our case, the Laplacians do have higher derivatives, it will be interesting to use the 
approach developed in \cite{gc} and analyse the spectral dimension. This issue will be taken up separately. Many types of 
modified diffusion equations were analysed in \cite{gc1} and shown to have solutions showing negative probability. For 
certain class of these modified equations, an approach to obtain solutions to diffusion equations involving higher 
derivatives having positive definite probability was discussed in\cite{gc1} and it has been applied to Horava-Lifschits 
gravity models as well as to anomalous diffusion, both in time and space in \cite{gc1}. It will be interesting to study 
the dimensional flow of $\kappa$-space-time, in terms of the Laplacians we used, using this approach and work along 
these lines are in progress.\\

{\noindent{\bf Acknowledgements}}: AV thank UGC, India, for support through BSR scheme.

\end{document}